\documentclass[aps,prd,floatfix,superscriptaddress,preprintnumbers,nofootinbib,twocolumn,longbibliography]{revtex4-2}
\pdfoutput=1

\usepackage[english]{babel}
\usepackage[dvipsnames]{xcolor} 
\usepackage{amsmath, amssymb, amsfonts, graphicx, mathrsfs, verbatim, float, slashed, bbm, xspace, enumerate, url, bbding, multirow, outlines, upgreek, braket, mhchem,tikz}

\usepackage{hyperref}
\hypersetup{
     colorlinks   = true,
     citecolor    = blue,
     urlcolor     = blue,
     linkcolor    = blue
}

\definecolor{lime}{HTML}{A6CE39}
\DeclareRobustCommand{\orcidicon}{%
	\begin{tikzpicture}
	\draw[lime, fill=lime] (0,0) 
	circle [radius=0.16] 
	node[white] {{\fontfamily{qag}\selectfont \tiny ID}};
	\draw[white, fill=white] (-0.0625,0.095) 
	circle [radius=0.007];
	\end{tikzpicture}
	\hspace{-3mm}
}

\foreach \x in {A, ..., Z}{%
	\expandafter\xdef\csname orcid\x\endcsname{\noexpand\href{https://orcid.org/\csname orcidauthor\x\endcsname}{\noexpand\orcidicon}}
}

\definecolor{ForestGreen}{rgb}{.34,.139,.34}

\begin{document}

\preprint{SLAC-PUB-17783}
\preprint{UCI-HEP-TR-2024-11}

\title{Loosely Bound Composite Dark Matter}

\author{\orcidA{}~Javier F. Acevedo~}
\thanks{\href{mailto:jfacev@slac.stanford.edu}{jfacev@slac.stanford.edu}}
\affiliation{Particle Theory Group, SLAC National Accelerator Laboratory, Stanford University, Stanford, CA, 94039, USA}
\affiliation{The Arthur B. McDonald Canadian Astroparticle Physics Research Institute, \\ Department of Physics, Engineering Physics, and Astronomy,\\ Queen's University, Kingston, ON, K7L 2S8, Canada}

\author{\orcidB{}~Yilda Boukhtouchen~}
\thanks{\href{mailto:yilda.boukhtouchen@queensu.ca}{yilda.boukhtouchen@queensu.ca}}
\affiliation{The Arthur B. McDonald Canadian Astroparticle Physics Research Institute, \\ Department of Physics, Engineering Physics, and Astronomy,\\ Queen's University, Kingston, ON, K7L 2S8, Canada}

\author{\orcidC{}~Joseph Bramante~}
\thanks{\href{mailto:joseph.bramante@queensu.ca}{joseph.bramante@queensu.ca}}
\affiliation{The Arthur B. McDonald Canadian Astroparticle Physics Research Institute, \\ Department of Physics, Engineering Physics, and Astronomy,\\ Queen's University, Kingston, ON, K7L 2S8, Canada}
\affiliation{Perimeter Institute for Theoretical Physics, Waterloo, Ontario, N2L 2Y5, Canada} 

\author{\\ \orcidD{}~Christopher Cappiello}
\thanks{\href{mailto:cappiello@wustl.edu}{cappiello@wustl.edu}}
\affiliation{The Arthur B. McDonald Canadian Astroparticle Physics Research Institute, \\ Department of Physics, Engineering Physics, and Astronomy,\\ Queen's University, Kingston, ON, K7L 2S8, Canada}
\affiliation{Perimeter Institute for Theoretical Physics, Waterloo, Ontario, N2L 2Y5, Canada}
\affiliation{Department of Physics, Washington University, St. Louis, MO, 63130, USA}

\author{\orcidE{}~Gopolang Mohlabeng}
\thanks{\href{mailto:gmohlabe@sfu.ca}{gmohlabe@sfu.ca}}
\affiliation{Department of Physics and Astronomy, University of California, Irvine, CA, 92697, USA}
\affiliation{Department of Physics, Simon Fraser University, Burnaby, BC, V5A 1S6, Canada}

\author{\orcidF{}~Narayani Tyagi}
\thanks{\href{mailto:narayani.tyagi@queensu.ca}{narayani.tyagi@queensu.ca}}
\affiliation{The Arthur B. McDonald Canadian Astroparticle Physics Research Institute, \\ Department of Physics, Engineering Physics, and Astronomy,\\ Queen's University, Kingston, ON, K7L 2S8, Canada}

\begin{abstract}
We investigate loosely bound composite states made of dark matter, where the binding energy for constituent particles is considerably less than the constituent mass. We focus on models of nuclear and molecular dark matter, where constituents are separated by length scales larger than the inverse constituent mass, just like nuclei and atoms in the Standard Model. The cosmology, structure, and interactions at underground experiments are described. We find that loosely bound composites can have a very large cross section for scattering with nuclei that scales with nucleon number like $\sim A^4$. For some couplings, these composites produce extremely soft ($\ll$ keV) individual atomic recoils while depositing a large amount of total recoil energy ($\gg$ keV) in a single passage through a detector, implying an interesting new class of signatures for low threshold direct detection.
\end{abstract}

\maketitle

\section{Introduction}
The nature of dark matter (DM) remains an intriguing puzzle for modern physics. It has long been appreciated that dark matter may be a composite state that is much larger in size or occupancy number than bound states like nuclei or atoms in the Standard Model (SM) \cite{Witten:1984rs,Farhi:1984qu,DeRujula:1984axn,Goodman:1984dc,Drukier:1986tm,Nussinov:1985xr,Bagnasco:1993st,Alves:2009nf,Kribs:2009fy,Lee:2013bua,Krnjaic:2014xza,Detmold:2014qqa,Jacobs:2014yca,Wise:2014ola,Hardy:2014mqa,Hardy:2015boa,Bramante:2018qbc,Gresham:2018anj,Bramante:2018tos,Ibe:2018juk,Grabowska:2018lnd,Coskuner:2018are,Bai:2018dxf,Digman:2019wdm,Bai:2019ogh,Bramante:2019yss,Bhoonah:2020dzs,Clark:2020mna,Gresham:2017zqi,Gresham:2017cvl,Wise:2014jva,Acevedo:2020avd,Acevedo:2021kly,Fedderke:2024hfy}. Prior composite DM studies have focused on composites with a few constituents or with many constituents that each have a sizable binding energy $E_B$ relative to the constituent particle mass outside the composite, $m_d$. In the case of asymmetric fermions bound together by a scalar field, composites with large constituent number usually have a binding energy that is a little less than the mass of an unbound constituent, $E_B \lesssim m_d$, making the mass per constituent $\overline{m}_d$ comparatively small, $\overline{m}_d \ll m_d \sim E_B$. These have been called ``saturated composites" \cite{Gresham:2017zqi,Gresham:2017cvl,Wise:2014jva,Acevedo:2020avd,Acevedo:2021kly}, and they have a high density of constituents tightly bound together. In this work, we will begin exploring less dense, loosely bound composites where the binding energy of a constituent is substantially smaller than the constituent mass so that $m_d \approx \overline{m}_d$. The cosmology and detection of such loosely bound composites differs from more tightly bound composites, and in particular we will find that the cross-section for coherent SM nuclear scattering on loosely bound composites can be very large, and lead to different regimes of interactions.  

In the visible sector, nuclei composed of protons and neutrons match this loosely bound description, since the nucleon mass is larger than the inter-constituent length scale, and much larger than the nuclear binding energy 
\begin{align*}
   m_{n} > \Lambda_{S} >  E_B^{\rm (nuclear)} , 
\end{align*} 
where $\Lambda_S \approx 100~{\rm MeV}$ is the confinement scale for the SU(3) color gauge force in the SM, which sets the inter-constituent spacing in nuclei, with nucleon binding energies $E_B^{\rm (nuclear)} \approx 10~{\rm MeV}$. Molecular composites in the SM, which are bound together by electrons with mass $m_e$ and fine structure constant $\alpha_e$, also follow this hierarchy of scales where the atomic mass is greater than the inverse Bohr radius $r_B^{-1} \approx \alpha_e m_e$ and much greater than the molecular binding energy $E_B^{\rm (molecular)} \approx \alpha_e^2 m_e$, 
\begin{align}
    m_n > r_B^{-1} > E_B^{\rm (molecular)}.
\end{align} 
Hereafter we will consider loosely bound composites that hew to these SM scalings, and also begin more widely surveying the loosely bound composite space, and in particular composites with binding energies smaller than their constituent mass by many orders of magnitude $E_B \ll m_d$. One prior example of such loosely bound composite dark matter is quirky composite dark matter \cite{Kribs:2009fy}, although that work considered a regime in which there are only a few constituents per composite. 

In contrast, we will see here that coherently enhanced scattering processes for loosely bound composites with many constituents markedly alter their characteristics and detection. Distinctive multiscatter interactions arise from large physical separation between constituents that can manifest in the dark sector as a confinement scale in a nuclear theory of dark matter, or as an inter-molecular scale akin to the Bohr radius for a dark molecular bound state. For both these composite models, we study couplings between dark constituents and SM nucleons in terms of a constituent-nucleon cross section $\sigma_{nd}$ and examine scattering signatures at experiments like DEAP-3600. We also study dark constituents coupling to electrons with a cross-section $\sigma_{ed}$, and for this we will find a unique regime of scattering where individual composites induce a very large number of low energy atomic recoils during a single passage through a dark matter detector.

The rest of this paper is as follows. In Section \ref{sec:cosmo} we outline models for loosely bound composites and discuss their cosmology and formation. Detection of loosely bound composites coupled to the SM via coupling to nucleons is detailed in Section \ref{sec:nuclear-scattering}, leading to predictions for number of nuclear scatters in a single passage through an experiment like DEAP-3600, presented in Section \ref{sec:deap}. Coupling of constituents to electrons is covered in Section \ref{sec:elec} along with a novel kind of composite-detector interaction, where a large amount of recoil energy is deposited over many low-recoil energy atomic scatters. We outline some calculations concerning future detection prospects in condensed matter-based detectors in Section \ref{sec:cond-mat-det}. Section \ref{sec:conc} concludes. Some formal details of loosely bound composite-nucleon scattering are presented in Appendix \ref{app:details}. Throughout we will be using natural units where $\hbar = c = 1$.

\section{Models and Cosmology for Loosely Bound Composites}
\label{sec:cosmo}

To begin an inquiry into loosely bound composite dark matter, we can consider a model where the dark matter is a ``dark nucleus," composed of dark nucleons held in stable bound states by either a dark confining force or dark electron. Such a model would arise from a dark sector composed of dark fermions charged under SU(N) (see $e.g.$ Refs.~\cite{Krnjaic:2014xza,Detmold:2014qqa}), which form dark nucleons with a confinement scale $\Lambda_D$ and corresponding radius $R_D = 1/\Lambda_D$, analogous to SM quantum chromodynamics. A coupling to a dark pion $\pi_d$, inducing an attractive force between nucleons, would induce the formation of larger ``dark nuclei" through $\pi_d$ emission: $d + d \rightarrow D_2 + \pi_d$. The subsequent formation of nuclei with larger constituent number has been previously studied \cite{Krnjaic:2014xza}, and depending on factors like bottlenecks, nuclei can grow to very large nuclear number.

Most of the above considerations can also be applied to a model of molecular dark matter, where in this case the dark atoms have a separation scale $\Lambda_D \approx \alpha_{de} m_{de}$ determined by the strength of a dark sector U(1) gauge coupling and the mass of a dark electron. In this case, the attractive force between dark constituents is provided by dark electrons, and the formation of dark molecules in the early universe would proceed through processes like $d+d \rightarrow D_2 + \gamma_d$, where $\gamma_d$ would be the photon field of the dark sector.

In the treatment that follows, we will estimate the number of constituents that would assemble to form composites, for a various binding energies and separation scales $\Lambda_D$. We will assume these composites are asymmetric, and have a relic abundance set by a dark sector asymmetry. We note that there are many viable models that obtain an asymmetric dark matter abundance, for fermion masses extending down to the keV range \cite{Petraki:2013wwa,Zurek:2013wia,Balan:2024cmq}. Assuming that loosely bound composites follow a ``liquid drop" model for their structure, the dark composite radius will scale with the dark constituent number $N_D$ as
\begin{equation}
    R_{D} \sim \frac{N_D^{1/3}}{\Lambda_D}
\end{equation}
In the early universe, the composites will grow in size through $N$-body binding processes, $D_N + D_N \rightarrow D_{2N}$. We can estimate the expected constituent number in a dark composite at the time when the binding rate drops below the Hubble rate, $\Gamma/H = \langle \sigma_{D_N} v_{D_N}\rangle n_{D_N} / H \sim 1$, where this expression equates the rate for composites to scatter with the Hubble rate. Re-expressing the interaction rate of the composites in terms of the constituent velocity, $v_{d}$, and cross-section, the $N_D$-constituent composite velocity will scale as $v_{D_N} = v_d N_D^{-1/2}$. 
We also assume the constituent cross section scales geometrically: $\sigma_{D_N} = 4 \pi N_D ^{2/3} /\Lambda_D^2$. Rewriting $n_{D_N} = n_d N_D^{-1}$, we find that the typical constituent number at the temperature of composite assembly $T_{ca}$ is,
\begin{equation}
    \begin{split}
        N_D &= \left (\frac{4 \pi n_d v_d }{\Lambda_D^2 H}\right)^{6/5} 
        = \left (\sqrt{\frac{18}{5}} \frac{\pi^{5/2} \sqrt{g^*_{ca}} T_{ca}^{3/2} T_r M_{pl}}{\zeta \overline{m}_d^{3/2} \Lambda_D^2} \right)^{6/5} \\
        &\simeq 2.5 \times 10^{13} \left(\frac{10}{g^*_{ca}}\right)^{3/5}\left(\frac{T_{ca}}{0.01 \text{ GeV}}\right)^{9/5}\left(\frac{10^{-5}}{\zeta}\right)^{6/5} \\
        &\hspace{15pt}\left(\frac{10 \text{ GeV}}{\overline{m}_d}\right)^{9/5}\left(\frac{\text{ GeV}}{\Lambda_D}\right)^{12/5}
    \end{split}
    \label{eq:Nd}
\end{equation}
where in the first equality, we have used the Friedmann equation, $3 H^2 M_{pl}^2 = g_{ca}^* \pi T^4/30$, the dark constituent velocity $v_d = \sqrt{T/\overline{m}_d}$, and estimated the dark constituent number density by $n_d = g_{r}^* \pi^2 T_{ca}^3 T_r/ 30 \zeta \overline{m}_d$, where we take $T_r \simeq 0.8$ eV, $g_{r}^* \simeq 3$ as the temperature and effective relativistic degrees-of-freedom  at matter-radiation equality \cite{Aghanim:2018eyx}. For cosmologies in which the dark matter abundance is diluted after composite assembly, $\zeta = s_{\rm before}/s_{\rm after}$ parametrizes the dilution factor due to, $e.g.$ a decaying field, or any other source of early universe entropy injection that depletes the final abundance of dark composites \cite{Bramante:2017obj}. The temperature of composite assembly $T_{ca}$ will be determined by the binding energy of the two-body state $\rm BE(2)$, and is approximately given by $T_{ca} \sim {\rm BE}(2)/10$. 

We next discuss the binding energy per constituent for loosely bound composites. In the case of molecular dark matter, we have already noted that the binding energy will be determined by the dark sector gauge coupling and dark electron mass. For loosely bound nuclear dark matter, we can consider a Yukawa potential describing the attractive interaction between dark nucleons, like in the SM. Then at large $N$, the binding energy per nucleon can be determined with the liquid drop model, which, in the SM is determined by Coulomb, degeneracy, and attractive strong binding terms.
In this case the binding energy per constituent for dark nuclei would be proportional to \cite{greiner2012nuclear}
\begin{equation}
    \frac{{\rm BE}(N_D)}{N_D} \propto a_V - a_S N_D^{-1/3} - a_C N_D^{2/3}~,
\end{equation}
where $a_V,a_S,a_C$ are standard liquid drop model coefficients with dimensions of energy. 
More generally, if we consider a Yukawa potential $\propto \exp{\left(-m_{\pi_d} r\right)}/r$ with range $m_{\pi_d}^{-1}$, then we would have
\begin{equation}
    \frac{{\rm BE}(N_D)}{N_D} = a_V'\frac{\Lambda_D^3}{\left(m_{\pi_d}\right)^2} - a_S'\frac{\Lambda_D^4}{\left(m_{\pi_d}\right)^3} N_D^{-1/3} - a_c'\Lambda_DN_D^{2/3}~,
\end{equation}
where have re-expressed the liquid drop coefficients in terms of $\Lambda_D$ and $m_{\pi_d}$, and $a_V',a_S', a_C'$ are unitless.
In the following treatment, for the sake of simplicity we will be assuming a simple model for our nuclear binding energies where
\begin{equation}
    \frac{{\rm BE}(N_D)}{N_D} \approx a_V' \Lambda_D~,
\end{equation}
which for $a_V' \lesssim 0.1$ and for negligible Coulomb interactions, will be a good approximation for a wide range of model space. We will allow the dark confinement scale to take on a wide range of values. Finally we note that in this work, we will consider $\Lambda_D \gg 1$ eV, so that confinement and composite formation occurs before matter-radiation equality, as discussed previously, $cf.$ Eq.~\eqref{eq:Nd}.

\section{Regimes of Composite-Nucleus Scattering}
\label{sec:nuclear-scattering}
\begin{figure}
    \includegraphics[width=\columnwidth]{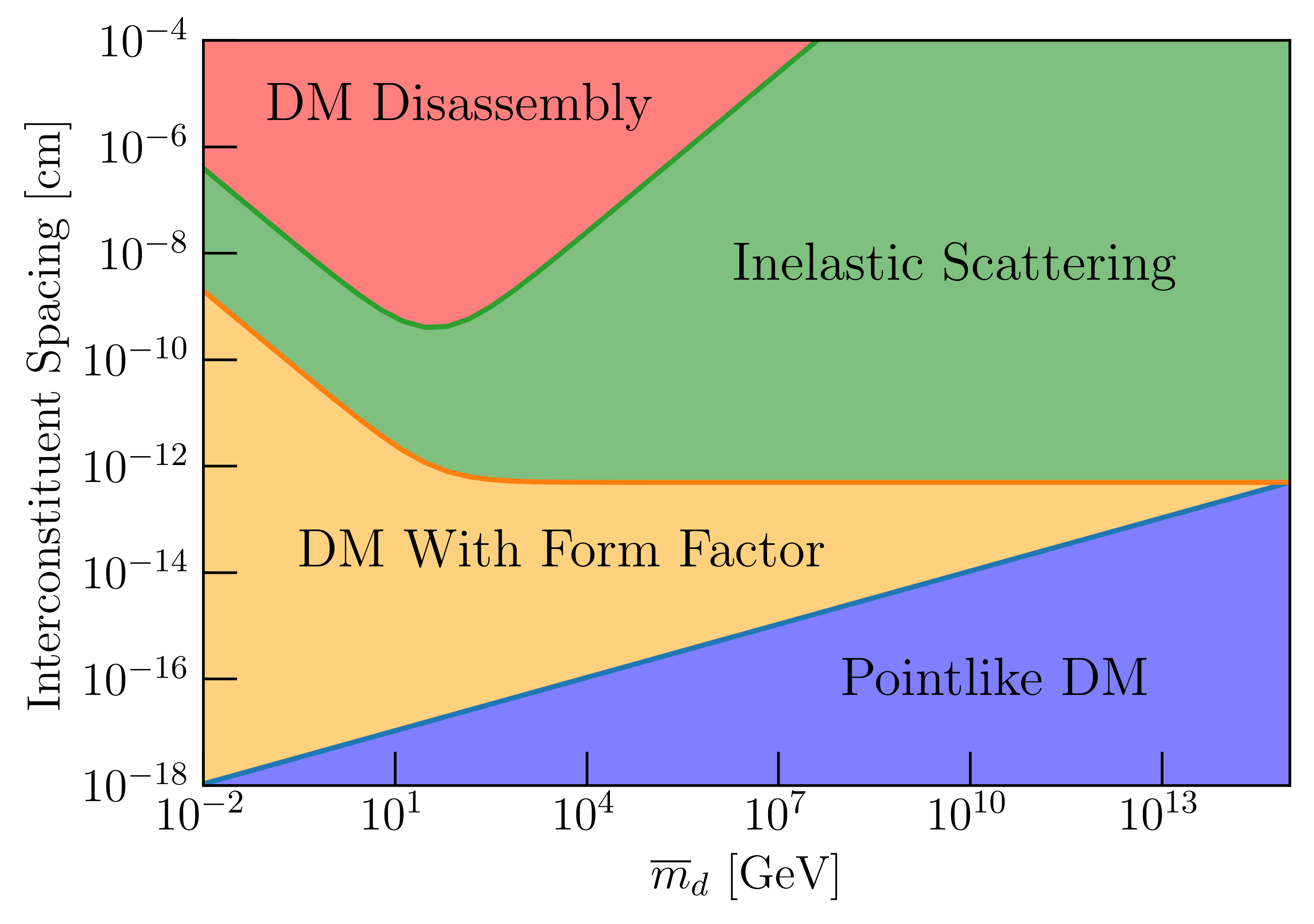}
    \caption{Different DM-nucleus scattering regimes for composite dark matter with total mass $M_D = N_D \overline{m}_d = 10^{15}$ GeV. The x-axis shows the constituent mass inside the composite $\overline{m}_d$, while the y-axis shows the inter-constituent spacing $\Lambda_D^{-1}$. Here the binding energy per constituent has been fixed to ${\Lambda_D}/{10}$. The momentum exchange between DM and nucleus has been fixed using a collisional velocity $v_D = 10^{-3}$. For the smallest inter-constituent spacing, nuclear scattering will proceed with the DM composite being pointlike. For larger spacing, a DM form factor is required, until at a large enough spacing determined by the nucleus-constituent reduced mass, nuclei will scatter with individual constituents through an inelastic exchange, and for small enough binding energies, through an exchange that removes constituents from the composite.}
    \label{fig:compositespace}
\end{figure}

In this section, we outline the different regimes for loosely bound composite dark matter scattering in a detector, parametrized by the constituent spacing scale $\Lambda_D$. When the size of the composite is small, either because $\Lambda_D$ is large or the number of constituents is small (or both), the composite behaves like a point particle. For slightly more loosely bound composites (intermediate $\Lambda_D$), the dark matter scatters elastically, but with the cross section suppressed by the form factor. For even more loosely bound composites, the momentum transfer is large enough for a nucleus to scatter with individual constituents, leading to incoherent scattering which can excite the composite (which we denote ``inelastic scattering"). And finally, for extremely loosely bound composites, the binding energy may be insufficient to hold the composite together following a collision, resulting in spallation of the DM composite. In Fig.~\ref{fig:compositespace}, we show the values of the in-medium constituent mass $\overline{m}_d$ and the interparticle spacing $\Lambda_D^{-1}$ that correspond to these different scattering regimes, for a composite state with a total mass $M_D = 10^{15} \ \rm GeV$.

\subsection{Large $\Lambda_D$ (Pointlike Dark Matter)}

In the case of arbitrarily large $\Lambda_D$, the radius of the composite is smaller than that of a nucleus, and smaller than any momentum transfer in the problem, meaning that the dark matter form factor is always equal to one. This is the point-like limit, where the scattering is always coherent across the dark composite. When the coupling to nuclei is small, the Born Approximation holds, and the composite DM-nucleon scattering cross-section $\sigma_{nD}$ is given by

\begin{equation}
    \frac{d \sigma_{nD}}{dE_R} = N_D^2 \left(\frac{\mu_{nD}}{\mu_{nd}}\right)^2 \frac{d \sigma_{nd}}{dE_R},
\end{equation}
where $\mu_{nD}$ is the composite-nucleon reduced mass, $\mu_{nd}$ is the constituent-nucleon reduced mass, and $E_R$ is the nuclear recoil energy. Note that the reduced mass ratio will reduce to $N_D$ as long as the mass of the composite and constituent are much larger than that of a nucleon, and is not dominated by binding energy. The phenomenology of this case is then identical to that of a weakly interacting massive particle with mass $M_D$.

If the coupling to nuclei is made too large, then the Born Approximation breaks down, and for a repulsive DM-nucleus interaction, the cross-section saturates to approximately the geometric size of the nucleus. The confinement scale $\Lambda_D$ becomes large enough for the DM to be considered point-like when the composite radius is small compared to the maximum allowed momentum transfer. For a heavy composite scattering with xenon, that maximum momentum transfer is about 240 MeV. So the dark matter is pointlike for $R_D \ll 1$ fm. For scattering with argon, the maximum momentum transfer is about 75 MeV, so we still find that the pointlike limit is $R_D \ll 1$ fm.

\subsection{Intermediate $\Lambda_D$ (Composite With Form Factor)}
When $\Lambda_D$ is larger than about 1 fm (for scattering with xenon), and the coupling to nucleons is small, the DM form factor begins to suppress the coherent scattering cross section. In this subsection we consider the case that $\Lambda_D$ is small enough that the composite is not point-like, but larger than the energy a nucleus can transfer to a constituent ($i.e.$ scattering with a nucleon cannot break apart the composite). The differential cross section is

\begin{equation}
    \frac{d\sigma_{nD}}{d E_R} = g^2 N_D^2 \left(\frac{\mu_{nD}}{\mu_{nd}}\right)^2\frac{d\sigma_{nd}}{dE_R}|F_D(E_R)|^2\,,
    \label{eq:coherent1}
\end{equation}
where
\begin{align}
    g^2 = \textrm{min}\left[1,\,\left(\frac{L_{A}}{R_D}\right)^3\right]^2\,
\end{align}
accounts for the overlap of the DM with the wavefunction lengthscale $L_A$ of the nucleus, which we take as the thermal de Broglie wavelength of the atom. When the radius of the composite is very large compared to the thermal de Broglie wavelength of the atom, this overlap factor would seem to heavily suppress the scattering rate, and if the composite were scattering with an individual atom, this would be the case. But when the dark matter passes through a medium, as in a detector or the crust of the Earth, the composite will overlap with many atoms, and a volume integral must be performed to compute the proper scattering rate. In this section, we are always in the regime where $R_D \lesssim L_A$, so for simplicity we neglect performing such a volume integral.

So long as $N_D$ is sufficiently large, this coherent scattering term, Eq.~\eqref{eq:coherent1} can dominate over the incoherent scattering cross section described in the next section. However, we see that for large enough momentum transfer, this coherent scattering cross section will become suppressed. How strongly it is suppressed depends on the choice of the dark form factor $F_D(q)$. The Helm form factor has an exponential falloff, but this is due to the nuclear skin depth factor. If we neglect this part of the Helm form factor, we get the square of a spherical Bessel function, which goes like $q^{-4}$ times sinusoidal resonances. We discuss some particulars of this in Appendix \ref{app:details}. Irrespective of form factor choice, once the coupling to nucleons are large enough, the dark matter form factor ceases to be a relevant quantity, since at such a large implied cross section between the SM nucleus and the composite, the Born Approximation for scattering on the composite fails. If the dark composite potential is extremely large, then a nucleon is not able to penetrate the dark matter composite to probe its internal structure. In that case, the total cross section again reaches the geometric limit of the composite's area. 

For small coupling to nucleons, $i.e.$ where the $N_D$ scaling still holds, the cutoff could be higher: the point where the composite gets large enough that the form factor suppression fully compensates for the factors of $N_D$ in the coherent cross section, and incoherent scattering ($i.e.$ excitation of the composite) takes over. 

\subsection{Small $\Lambda_D$ (Incoherent Scattering: Inelastic / Disassembly)}

When $\Lambda_D$ is small, and the coupling to nucleons is small, coherent scattering is totally shut off, and the scattering is between a nucleus and a constituent.  This may lead to excitation of the constituent inside the composite, or, if the composite is loosely bound enough, the collisions could knock constituents out of the composite, which could lead to the disintegration of the dark matter state. 
The constituent-nucleus cross section in this case carries the standard $\mu^2A^2$ dependence of point-like dark matter scattering with a nucleus, and the composite-nucleon cross section is just
\begin{equation}
    \frac{d\sigma_{nD}}{dE_R} = g\, N_D\frac{d\sigma_{nd}}{dE_R}S_{D}(q)\,.
\end{equation}
The function $S_D(q)$ encodes the kinematics of a constituent being either fully liberated from the composite, or upscattered into an excited state, the latter of which depends on the energy level structure of the composite and on which energy levels are Pauli blocked. In the limit of an extremely weakly bound composite, any constituent that is struck by a SM particle is fully liberated from the composite, and $S_D(q) = 1$. We thus assume here that $S_D(q) = 1$, and leave the computation of $S_D(q)$ for more strongly bound composites to future work.

In the case of large coupling, the constituent-nucleus cross section saturates to the geometric size of the nucleus. The total cross section to scatter with the constituent is Min[$4\pi R_D^2$,$N_D \times 4\pi R_A^2$], $i.e.$ either the size of the nucleus times number of targets, or just the size of the composite, depending on how diffuse the constituents are in the composite.

\section{A Concrete Example: DEAP-3600 Search for Supermassive Dark Matter}
\label{sec:deap}

As a concrete example, we examine the limits on supermassive dark matter recently reported by the DEAP collaboration~\cite{DEAPCollaboration:2021raj}. This analysis considered two models for the DM-nucleus interaction. In the first model, the DM-nucleus scattering is equal to the DM-nucleon cross section times the nuclear form factor:
\begin{equation}
    \frac{\textrm{d}\sigma_{Ad}}{\textrm{d}E_R} = \frac{\textrm{d}\sigma_{nd}}{\textrm{d}E_R}|F_A(q)|^2\,.
\end{equation}
This model is described as a scenario in which the nucleus is opaque to dark matter.
The second model includes the traditional $A^4$ scaling seen for heavy dark matter with spin-independent interactions
\begin{equation}\label{eq:coherent}
    \frac{\textrm{d}\sigma_{Ad}}{\textrm{d}E_R} \simeq A^4\frac{\textrm{d}\sigma_{nd}}{\textrm{d}E_R}|F_A(q)|^2\,,
\end{equation}
where the approximate equality becomes an equality when $m_{d} \gg m_A$. Hereafter, we will examine the limits shown for this second model, and show how the scaling of Eq.~\eqref{eq:coherent} can be obtained from a simple model of loosely bound composite dark matter.

\begin{figure}
\includegraphics[width=0.5\textwidth]{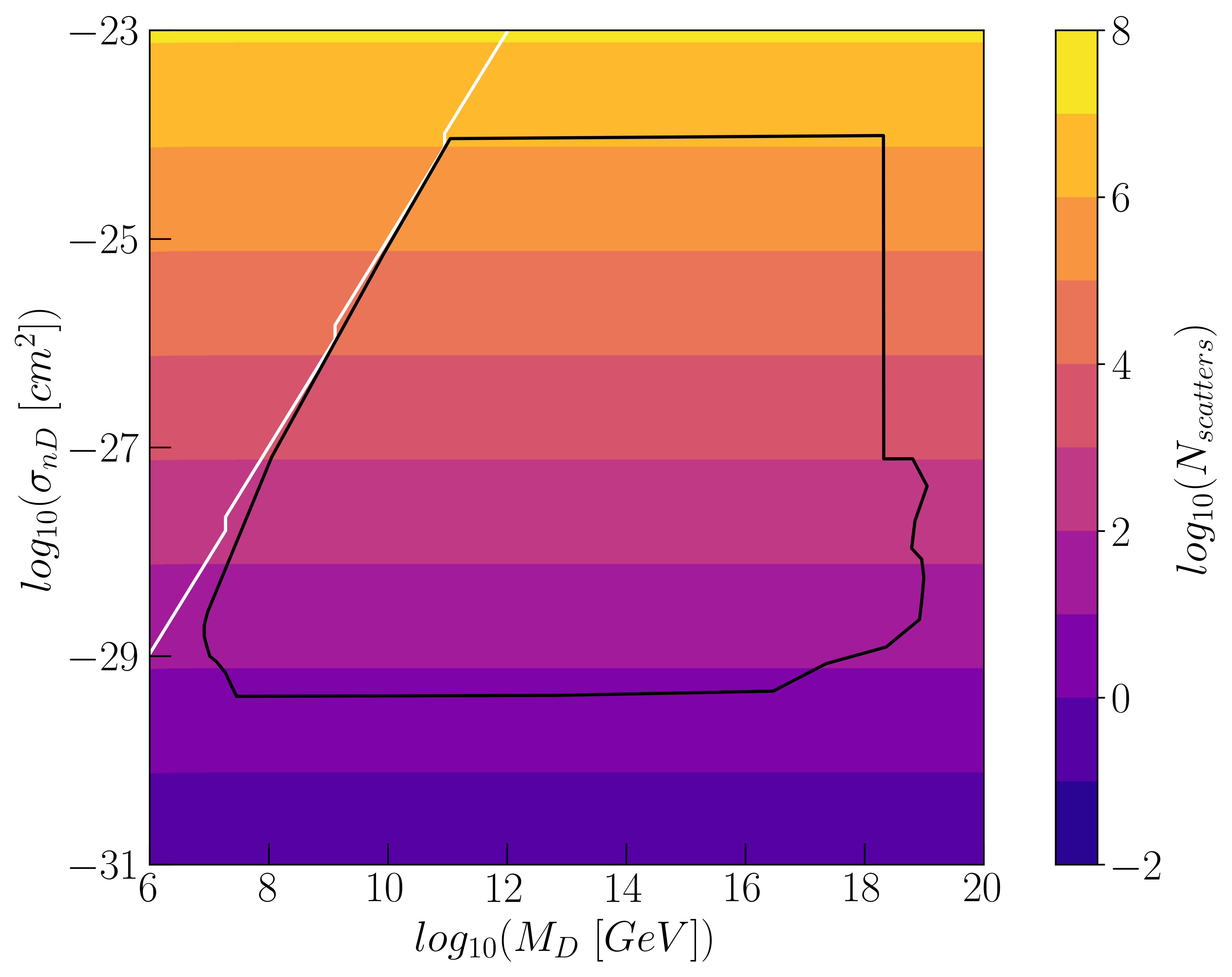}

\caption{\label{fig:deap} Log of the number of DM-nucleus collisions in the DEAP-3600 detector, using the second model assumed by the DEAP collaboration, in which the DM-nuclear scattering cross-section is assumed to scale with nucleon number as $A^4$. The black bounded region is the DEAP-3600 exclusion region, reproduced from Ref.~\cite{DEAPCollaboration:2021raj}. The bottom of the exclusion region is where the number of scatters is too small to constrain. The top of the exclusion region is where scattering is so frequent that the DEAP analysis breaks down due to individual light pulses merging together. On the right of the exclusion region, the DM flux is too low to constrain, while on the left, attenuation in the Earth causes the DM to lose energy before reaching the detector. The region above the white line represents parameter space where attenuation due to the Earth's overburden causes the dark matter to lose 99\% of its kinetic energy before reaching the DEAP-3600 detector; we see that the attenuation approximation used in this work closely matches the DEAP-3600 calculation.}
\end{figure}

Fig.~\ref{fig:deap} shows the parameter space probed by the DEAP-3600 detector. The black bounded region is the exclusion region reported in Ref.~\cite{DEAPCollaboration:2021raj}. The bottom of the exclusion region is where the number of collisions is too small to constrain with the DEAP analysis, while the flat top of the exclusion region is where there are so many collisions that the light pulses produced by individual collisions merge together, causing the analysis procedure to break down. For masses above about $10^{19}$ GeV, the DM flux is too small to constrain. The diagonal ceiling is due to DM being slowed in the Earth before reaching the detector.
The color scale is the base-10 log of the number of collisions that a DM particle produces within the DEAP-3600 detector. We see that the bottom and top of the DEAP region correspond to O(1) and O($10^6$) collisions, respectively. We account for attenuation in the Earth by setting the scattering rate to zero if the DM has lost 99\% of its kinetic energy before reaching the detector. We model the Earth overburden as composed purely of silicon, as the target with which scattering will dominate. We estimate the density of the overburden as 2.7 g/cm$^3$, as appropriate for the continental crust of the Earth, and set its depth as 2 km. 

This is not a perfect way of modeling attenuation, but it is a simple approximation that matches the ceiling reported by DEAP quite well.

In the remainder of this section, we consider the three distinct regimes for composite DM discussed in the previous section, and in each regime we ask whether it is possible to reproduce the event rates shown in Fig.~\ref{fig:deap}.

\subsection{Pointlike Dark Matter}

We first consider a repulsive DM-nucleon interaction, in the case where the DM is a composite with $N_D = 10^4$ and $\Lambda_D = 100$ GeV. The binding energy here is large enough that the dark matter is essentially pointlike: that is, the geometric size of the DM is smaller than any other distance scales in the problem, namely the size of an argon nucleus and the inverse momentum transfer associated with a collision. As discussed in Ref.~\cite{Digman:2019wdm}, for pointlike DM with a repulsive interaction, the scattering cross section cannot be larger than approximately the geometric size of the nucleus. 

In this case, the DM form factor, and indeed the fact that the DM is composite at all, is negligible, and any scattering with the DM is coherent across the whole composite. The DM-nucleus differential cross-section, derived in Appendix \ref{app:details} is

\begin{align}
    \frac{d\sigma_{AD}}{dE_R} = \left(\frac{\mu_{AD}}{\mu_{nd}}\right)^2A^2N_D^2 g^2 \frac{d\sigma_{nd}}{dE_R}F_A^2(q)F_D^2(q)\,,
\end{align}
where the composite form factor $F_D(q)$ is defined in the same way as the usual nuclear form factor.

\begin{figure}
\includegraphics[width=0.5\textwidth]{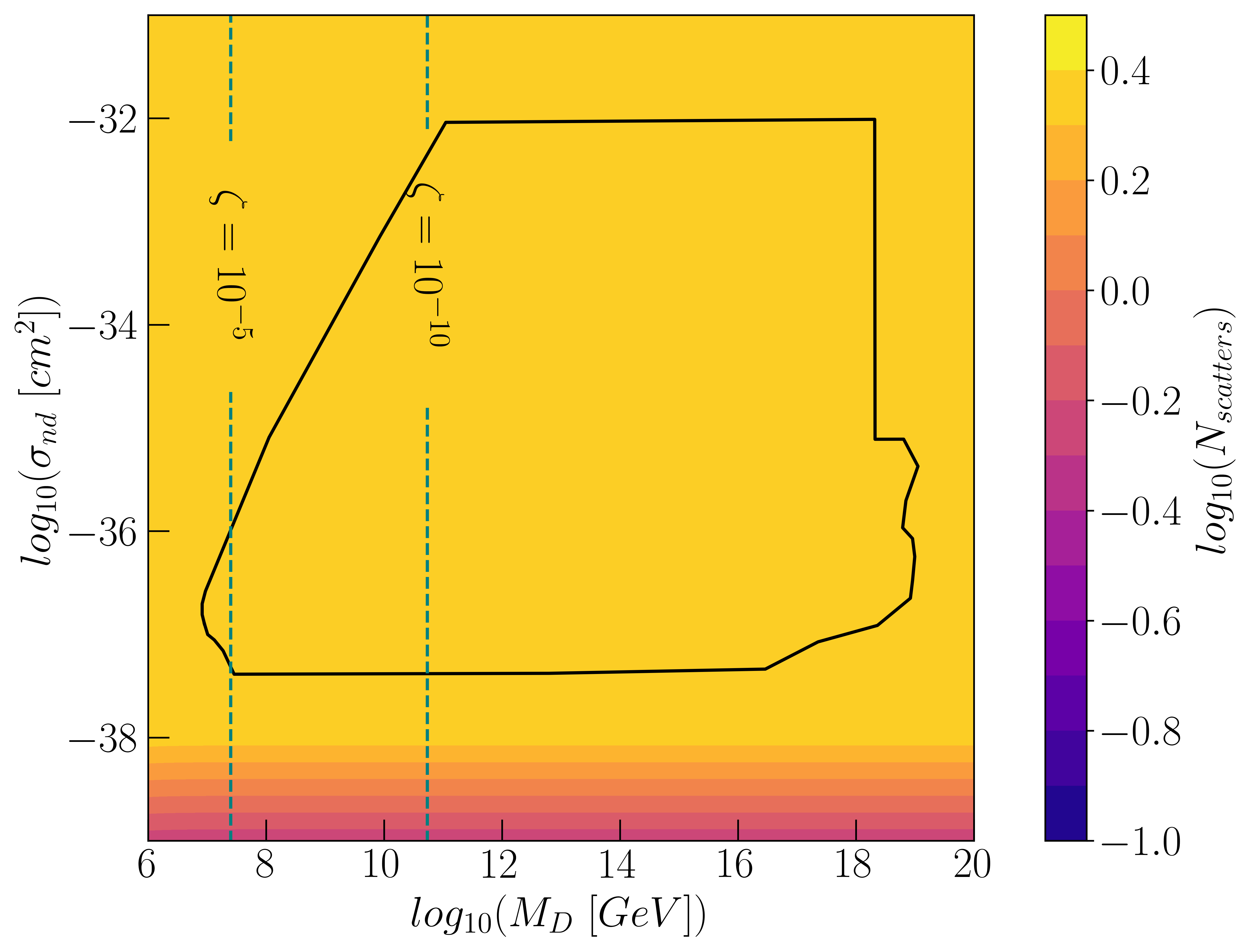}

\caption{\label{fig:lambda1} Log of the number of DM-nucleus collisions in the DEAP-3600 detector, using a model of pointlike DM with $N_D=10^4$, $\Lambda_D = 100$ GeV. In this case, DM-nucleus scattering is coherent across the entire composite, and so $\sigma_{nd} \propto \sigma_{nD}/N_D^2$, hence the DEAP-3600 region being shifted downward by exactly 8 orders of magnitude. The contour lines show the value of the dilution factor $\zeta$ that would give rise to composite states with these masses, according to Eq.~\eqref{eq:Nd}. The attenuation region lies above the parameter space plotted in this figure.}
\end{figure}

Fig.~\ref{fig:lambda1} shows the number of times a pointlike DM particle scatters with argon when passing through the DEAP-3600 detector. We see that above a cross section of about $10^{-30}$ cm$^2$, the event rate in the detector becomes constant. This is due to the aforementioned geometric limit for pointlike DM. We also see that the effect of attenuation in the Earth becomes largely negligible, because slowing the DM significantly in the Earth would require a cross-section larger than the geometric limit. Note that for smaller cross-sections, where the geometric limit is not relevant, the number of collisions is approximately equal to the number shown in Fig.~\ref{fig:deap}.

\subsection{Composite with Form Factor}

We can get around this geometric limit by making the composite DM geometrically large. When the DM is more physically extended than the nucleus, the geometric limit for the cross-section becomes the geometric size of the DM. However, when the DM is geometrically large, its form factor must be considered alongside the form factor of the nucleus, and in fact produces more suppression than the nuclear form factor.

\begin{figure}
\centering
\includegraphics[width=0.5\textwidth]{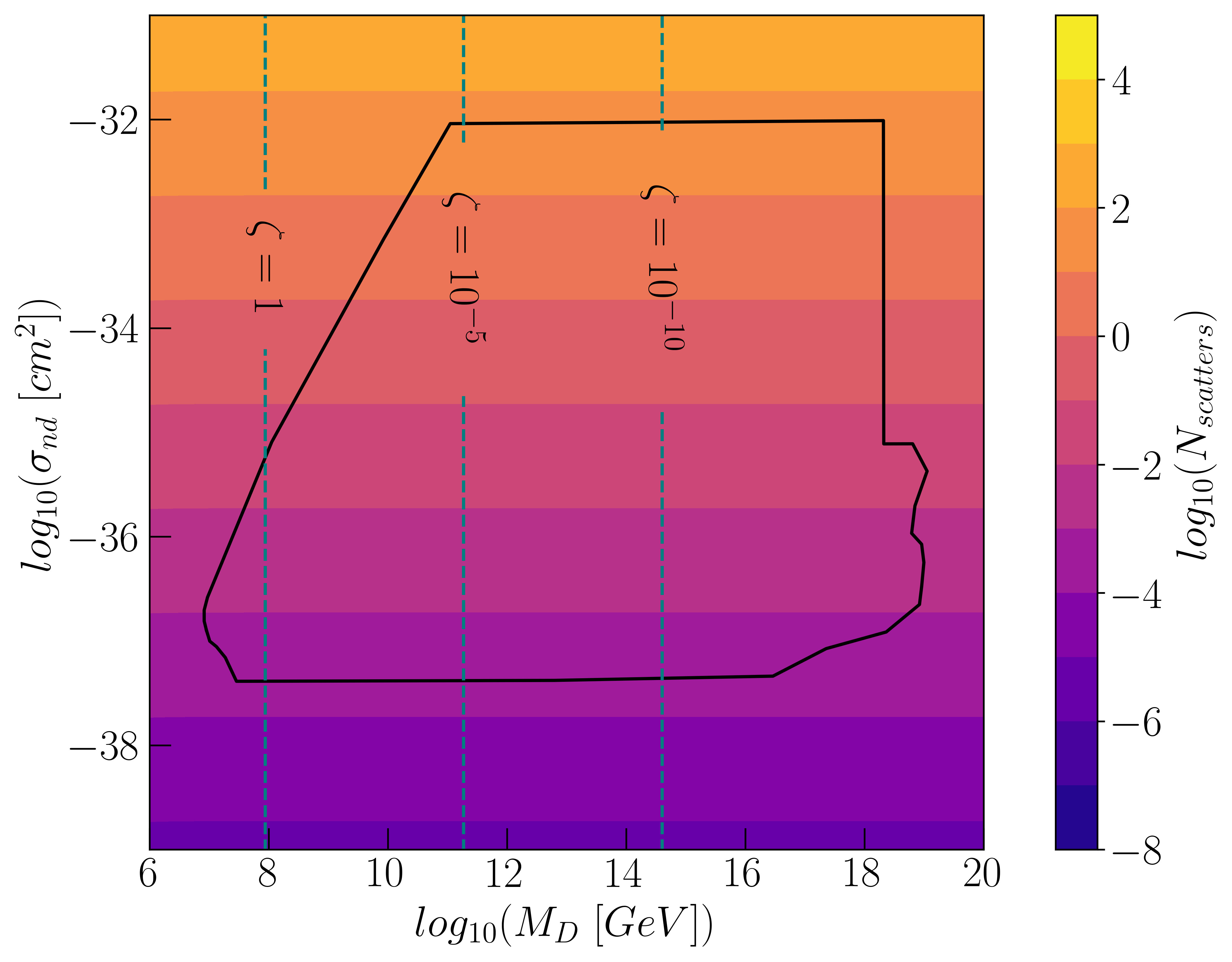}
\caption{\label{fig:lambdaminus2} Log of the number of DM-nucleus collisions in the DEAP-3600 detector, this time setting $N_D=10^4$, $\Lambda_D = 10$ MeV. The composite is more geometrically extended than the nucleus, so that the geometric limit is set by the DM composite size, and the cross-section must take into account the DM form factor. Here $\sigma_{nd} \propto \sigma_{nD}/N_D^2$, hence the DEAP-3600 region being shifted downward by exactly 8 orders of magnitude. The contour lines show the value of the dilution factor $\zeta$ that would give rise to composite states with these masses, according to Eq.~\eqref{eq:Nd}. The attenuation region lies above the parameter space plotted in this region.}
\end{figure}

Fig.~\ref{fig:lambdaminus2} shows the number of collisions in the detector for a geometrically extended DM composite, with the same number of constituents as before, but now $\Lambda_D = 10$ MeV. In this case, because the geometric limit is much larger, the number of collisions in the detector is no longer limited to $\sim$1. However, the presence of the DM form factor suppresses the number of collisions, so that at the bottom of the DEAP region, the number of collisions is $\ll 1$. We also see that attenuation is largely negligible, because the form factor not only suppresses the number of collisions, but also weights the scattering toward low energy transfer. 

Increasing the value of $\Lambda_D$ reduces the form factor suppression, but makes the geometric limit more of a problem. Decreasing the value of $\Lambda_D$ raises the geometric limit, but also makes the form factor suppression more severe. In general, we find that the scaling used by DEAP is not obtained via coherent scattering with an extended composite.

\subsection{Incoherent Scattering (Inelastic / Disassembly)}

On the other hand, we can consider the case of maximally incoherent scattering, where a nucleus sees the DM as a loosely bound collection of individual constituents. This is the case when $\Lambda_D$ is small compared to the typical momentum transfer between a nucleus and a constituent. The scattering is then described by the following equation, which is also derived in Appendix \ref{app:details}

\begin{equation}
    \frac{d\sigma_{AD}}{dE_R} = g\left(\frac{\mu_{Ad}}{\mu_{nd}}\right)^2A^2N_D\frac{d\sigma_{nd}}{dE_R}|F_A(q)|^2S_{D}(q)\,,
\end{equation}
where $S_{D}$ is a structure function for the composite accounting for the binding energy of constituents. We can see that as long as $m_d \gg m_A$, this cross section scales approximately with $A^4|F_A(q)|^2$. In what follows, we set $S_D(q) = 1$, which is appropriate for a loosely bound composite, where the binding energy has negligible effect on the scattering.

\begin{figure}
\centering
\centerline{\includegraphics[width=0.5\textwidth]{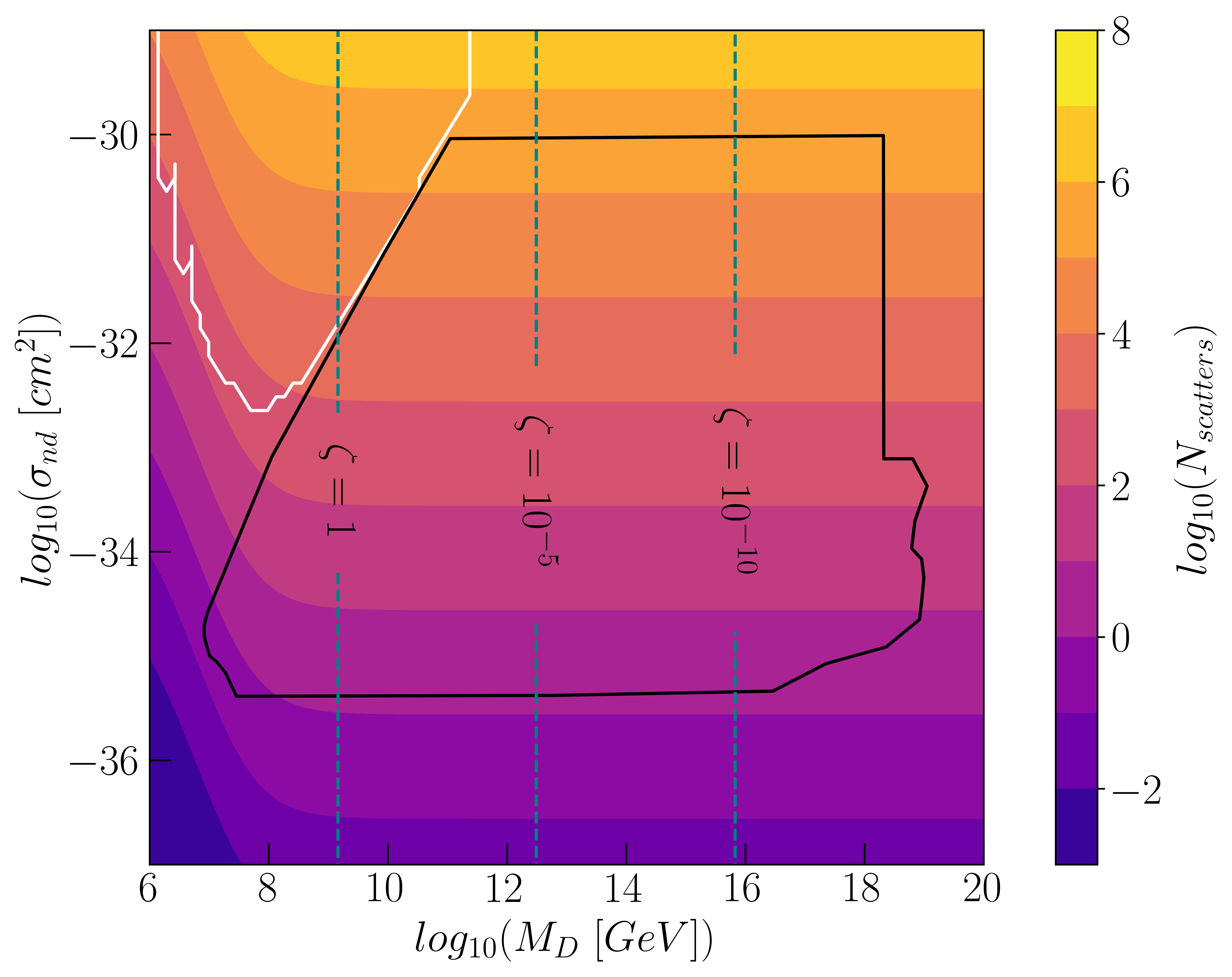}}

\caption{\label{fig:incoherent} Same as above, but for incoherent scattering with $N_D = 10^6$ and $\Lambda_D = 1$ MeV. In this case, scattering occurs between the nucleus and individual constituents in the DM composite. The constituent-nucleus scattering cross-section  is $\sigma_{nd} \propto \sigma_{nD}/N_D$, hence the DEAP-3600 detection region is shifted downward in number of scatters by around 6 orders of magnitude. The contour lines show the value of the dilution factor $\zeta$ that would give rise to composite states with these masses, according to Eq.~\eqref{eq:Nd}. The region above the white line represents parameter space where attenuation due to the Earth's overburden causes the dark matter to lose 99\% of its kinetic energy before reaching the DEAP-3600 detector.}
\end{figure}

Figure~\ref{fig:incoherent} shows the result for incoherent scattering, where we set $N_D = 10^6$ and $\Lambda_D = 1$ MeV. We see that for almost all of the DEAP-3600 parameter space, the event rate shown here approximately matches the event rate shown in Fig.~\ref{fig:deap}. The only difference comes at the lowest masses, and is due to the reduced mass of the constituent-nucleus system, $i.e.$, the fact that the above requirement that $m_d \gg m_A$ no longer holds.

\section{Composite-Atom Scattering}
\label{sec:elec}

In this section, we outline the dynamics for loosely bound composite dark matter interactions in the form of atomic recoils, as would be expected for dark matter coupled to an electron bound to an atom. In the low-momentum kinematic regime we will consider, the scattered electron remains in its original orbital instead of up-scattering to an excited state or ionizing the atom, so that the whole atom itself recoils \cite{Kopp_2009}. 

It is also possible to search for composite dark matter through ionizations or electron excitations, see $e.g.$ Ref.~\cite{Acevedo:2021kly}. In this paper, we will focus on the regime of coherent composite-atomic scattering. In this context, loosely bound composites present a novel dynamic, where many soft scatters are induced by dark matter during a single transit through a detector. The large number of constituents expected in loosely bound composites, as well as their large size, cause a large enhancement to the incoherent constituent-atom scattering. Thus, these composites would leave a signature of a potentially large number of expected scatters, each of which imparts very little energy. This is a distinct and interesting signature that may be sought out by future low-threshold detection experiments. 

\subsection{WIMP-Atom Scattering}

First we outline the standard WIMP-atom scattering. The differential cross-section for an atom with mass number A is
\begin{equation}
    \frac{d\sigma_{A d}}{d E_R} = \sum_{n, l} \frac{d\sigma_{e d}}{dE_R} | f_{n,l}(q)|^2 |F_{\phi} (q)|^2
    \label{eq:WAS}
\end{equation}
where $\sigma_{ed}$ is the reference cross-section for dark matter-electron scattering~\cite{PhysRevD.85.076007}, 
\begin{equation}
    \sigma_{ed} = \frac{\mu_{e d}^2}{16 \pi m_d^2 m_e^2} \overline{|\mathcal{M}_{e d}(q)|^2}|_{q^2 = \alpha^2 m_e^2}
    \label{eq:elecdm}
\end{equation}
such that the form factor $F_{\phi}(q)$ contains the momentum-dependence of the scatter, and in general, 
\begin{equation}
    \overline{|\mathcal{M}_{e d}(q)|^2}  = \overline{|\mathcal{M}_{e d}(q)|^2}|_{q^2 = \alpha^2 m_e^2} \times |F_{\phi}(q)|^2~.
\end{equation}
The form factor, assuming a fermionic constitutent and scalar mediator $\phi$, can be simplified in terms of $q$ as
\begin{equation}
    F_{\phi}(q) = \frac{\alpha^2 m_e^2 + m_{\phi}^2}{q^2 + m_\phi^2}~,
\end{equation}
which reduces to $F_{\phi}(q) = 1$ in the heavy mediator limit ($m_{\phi} \gg q$) and $F_{\phi}(q) = {\alpha^2 m_e^2}/{q^2}$ for light mediators ($m_{\phi} \ll q$). 

The atomic form factor $f_{n,l}(q)$ depends on the radial atomic wavefunctions $R_{n,l}$ \cite{Kopp_2009},

\begin{equation}
    \begin{split}
        f_{n, l}(q) &= \sum_m \braket{nlm| e^{i (\mathbf{k} - \mathbf{k'})\mathbf{x}}|nlm} \\
        &= (2 l + 1) \int dr \, r^2 |R_{nl}|^2 \frac{\sin{q r}}{q r}~,\\
    \end{split}    
\end{equation}
where the radial wavefunctions can be approximated through a linear combination of Slater-type atomic orbitals $S_{j,l}$. The wavefunctions are thus written as $R_{n, l} = \sum_j S_{jl} C_{jln}$, where $S_{jl}$ is a Slater-type orbital, $C_{jln}$ are coefficients, and the sum is over the linear basis set ${S_{j,l}}$ defined for the azimuthal quantum number $j$ \cite{Bunge1993}. At low momentum transfer ($q < 100 \ \rm keV$), $f_{n, l}(q)$ gives a $Z^2$ enhancement to the cross-section due to the number of available electron targets. At higher $q$, where the electron is more likely to be excited to a higher energy level, it strongly suppresses elastic atomic scattering.

\begin{figure*}[t!]
    \centering
    \includegraphics[width=\textwidth]{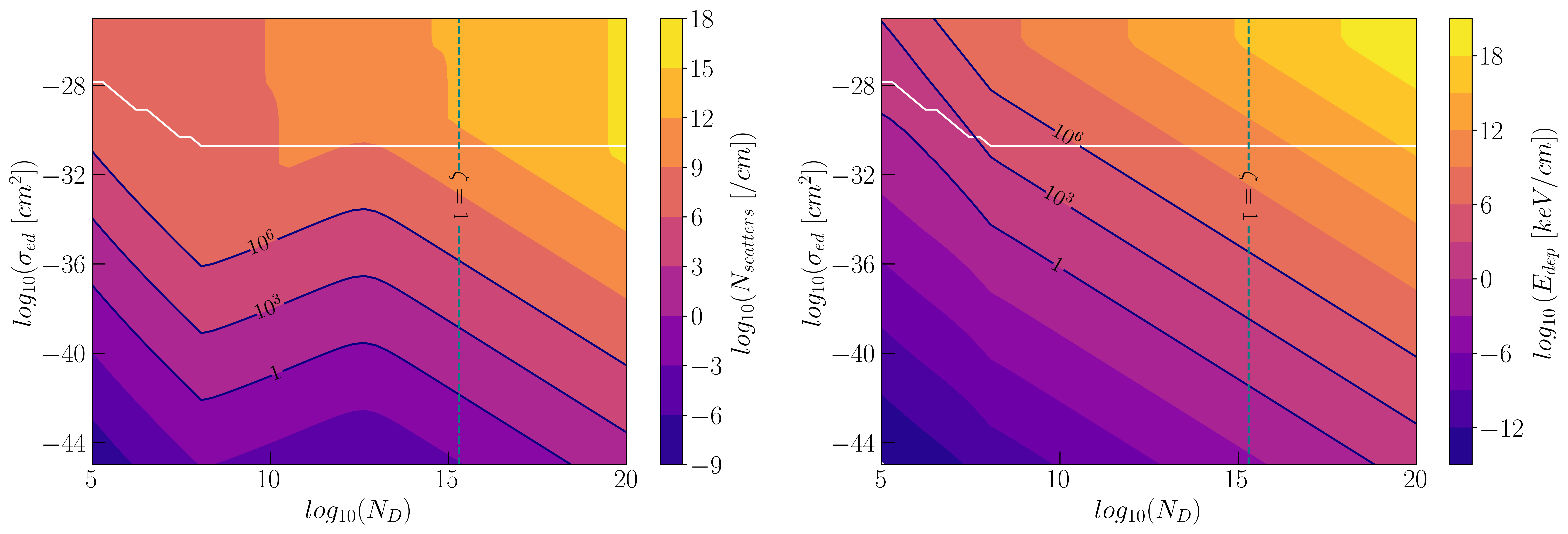}
    \caption{Number of scatters and energy deposited in a low threshold argon detector, for a composite dark matter model where constituents interact via a constituent-electron cross-section defined in Eq.~\eqref{eq:elecdm}. The region above the white line represents parameter space where attenuation due to the Earth's atmospheric overburden causes the dark matter to lose 99\% of its kinetic energy before reaching the detector. Here we have estimated this atmospheric column density as $\sim 10^3$ g/cm$^2$ of nitrogen \cite{Cappiello:2020lbk,Bramante:2022pmn}. \textit{Left}: Log of the number of DM-atom collisions over one cm in a liquid argon detector, using a model of composite dark matter with $\Lambda_D = 1$ MeV and $\overline{m}_{d} = 10$ MeV, implying a total composite mass $M_D = \overline{m}_d N_D$. \textit{Right}: total energy deposition in one centimeter for the same target.}
    \label{fig:AtomScatter}
\end{figure*}

\subsection{Regimes of Composite-Atom Scattering}

We now discuss composite dark matter scattering with an atom, once again in the three different regimes set by the confinement scale $\Lambda_D$.

\subsubsection{Large and Intermediate $\Lambda_D$: Coherent Scattering}
We first consider a composite scattering coherently with the target atom, which is appropriate when the composite size is smaller than the atomic size and the inverse momentum transfer of the problem. The coherent scattering cross-section is determined from Eq.~\eqref{eq:WAS},
\begin{equation}
    \begin{split}
    \frac{d\sigma_{A D}}{d E_R} &=g^2 N_D^2 \left(\frac{\mu_{e D}}{\mu_{e d}}\right)^2 \frac{d\sigma_{A d}}{dE_R}  |F_{D}(q)|^2~, \\
    \end{split}
    \label{atom-coh}
\end{equation}
where we include form factors to describe the spatial extent of the composite state, $F_{D}(q)$ and $g(L_{A})$. As in the nuclear scattering case, $F_{D}(q)$ is a composite form factor analogous to the nuclear form factor, and $g$ accounts for the limited overlap of the atomic wavefunction with the composite. In this case, we estimate the size of the atom as its Bohr radius, rather than de Broglie wavelength, when calculating $g$.

For a large $\Lambda_D$, the dark matter composite is essentially pointlike, and the scattering is just like WIMP-atom scattering. When the composite size becomes larger than the atom and inverse momentum transfer, $F_{D}(q)$  begin to suppress the scattering rate at large $q$, and there is a further suppression due to the partial overlap encompassed in $g$. Finally, the transition form factor $| f_{n,l}(q)|^2$ provides an enhancement at low momentum transfer ($q < 100 $ keV), and a strong suppression for higher $q$.

\subsubsection{Small $\Lambda_D$: Incoherent Scattering}

In the incoherent scattering limit, the scattering will be with individual constituents, and
\begin{equation}
    \begin{split}
    \frac{d\sigma_{AD}}{d E_R} &= g N_D \frac{d\sigma_{A d}}{dE_R} |S_D(q)|^2 \\
    \end{split}
    \label{eq:atom-inc}
\end{equation}
where as before, $S_D(q) = 1$ for a loosely bound composite: so long as the binding energies are well below the constituent recoil energies, we treat the constituents as a cloud of unbound dark matter particles. In the regime where $R_D \gg R_{A}$, we must perform a volume integral to calculate the total scattering rate, to account for the fact that the composite overlaps with many atoms. We estimate that the number density of constituents in the composite is constant, and thus the volumetric integral induces an additional factor of $R_D^3 \sim N_D \Lambda_D$ to the scattering rate.

\subsection{Searching for Composite-Atom Scattering with Low-Threshold Detectors}

So far we have seen that composite-atom scattering can have an $N_D$ enhancement to the cross-section when the composite is larger than an atom, so we can expect there will be multi-scattering on atoms by sufficiently large, loosely bound composite dark matter states. Indeed, we will find that a single composite state traversing a detector can cause a large number of scatters, each imparting a small amount of energy to the target nucleus, in certain experiments. Here we will focus on low-threshold DM searches aimed at sub-GeV DM searches. In this section, we outline what this multi-scattering atomic recoil signature would be like in a low-threshold liquid argon experiment, akin to the Scintillating Bubble Chamber \cite{SBC:2021yal}.

In Fig. \ref{fig:AtomScatter}, we show the base-10 log of the number of atomic scatters expected from one composite traversing a centimeter of argon, as well as the total energy deposition in the material. We have assumed a liquid argon density of $1.4$ g/cm$^3$. For our dark matter model parameters, we have fixed the length scale $\Lambda_D = 1 \ \rm MeV$ and fixed the constituent mass to be an of magnitude larger than $\Lambda_D$, so that we are in a  loosely bound regime, while varying the constituent number $N_D$ and constituent-electron cross-section $\sigma_{ed}$. We also assume a heavy mediator ($F_\phi(q)$ = 1). As in previous sections, we overlay on each plot some contour lines showing the value of the dilution factor $\zeta$ that would give rise to such composite states, according to Eq.~\eqref{eq:Nd}. 

In the leftmost region of Fig.~\ref{fig:AtomScatter}, where the composites are smallest, coherent scattering is the dominant scattering regime. At $N_D \approx 10^{8}$, the radius of the composite begins to exceed the radius of an argon atom, $R_D > R_{Ar}$, and coherent scattering is increasingly suppressed. At $N_D \approx 10^{13}$, the incoherent scattering rate becomes dominant and we see a large enhancement to the scattering rate as $N_D$ increases. In the right panel of Figure \ref{fig:AtomScatter} we see that a substantial amount of energy can be deposited over a centimeter of liquid argon, even if individual recoil energies for constituents are rather small, where in this case for $\bar{m}_d = 10$ MeV, constituent-atom scattering on argon will result in individual recoil energies $\sim 1$ eV - $1$ keV. It will be useful in further work to determine the response of a low-threshold liquid argon detector to such a trail of low-energy depositions.


\section{Searches with Condensed Matter Systems}
\label{sec:cond-mat-det}
The phenomenology and detection prospects of loosely bound composite states in low-threshold condensed matter-based detectors will be the subject of future work. However, for completeness, we provide here some details on how to estimate scattering and energy deposition rates in these systems (see $e.g.$ Ref.~\cite{Kahn:2021ttr}). We will focus on electrophilic couplings and solid-state detectors, such as conventional semiconductors and metals. In this scenario, the rate of excitations produced by a constituent within a composite state moving at velocity $v_D$, is expressed as 
\begin{equation}
    \Gamma(v_D) = \frac{\pi \sigma_{ed}}{\mu^2_{ed}} \int \frac{d^3 \mathbf{q}}{(2\pi)^3} \, |F_{\phi}(q)|^2 S(\mathbf{q},\omega)
    \label{eq:scat_rate_solid}
\end{equation}
where the energy transfer $\omega = \mathbf{q} \cdot \mathbf{v}_D - q^2/2m_d$ and the dynamic structure function 
    \begin{equation}
    S(\mathbf{q},\omega) = \frac{q^2}{2 \pi \alpha_e} \, {\rm Im}\left[-\frac{1}{\varepsilon(\mathbf{q},\omega)}\right]
    \label{eq:elf}
\end{equation}
is proportional to the imaginary part of the inverse dielectric function $\varepsilon$ in Fourier space through the Fluctuation-Dissipation Theorem. This can be computed from a model that accounts for the microphysics of the material, or experimentally measured since it is directly related to the response of the system to an external electron. The exact functional form sensitively depends on the kinematic regime, $i.e.$ the range of values of the momentum transfer $\mathbf{q}$ and energy transfer $\omega$. Note that the energy deposition rate can be estimated by multiplying by an additional factor $\omega$ in the integration in Eq.~\eqref{eq:scat_rate_solid}.

For composite states with $\overline{m}_d \ll 100 \ \rm MeV$, the accessible range of energy/momentum transfer becomes smaller than the Fermi momentum and electron plasma frequency of most common materials, and screening and many-body effects become relevant. In other words, the dynamic structure function deviates significantly from the single nuclear recoil branch. For this reason, the detection prospects of loosely bound composite states, for a variety of solid-state detectors, will be explored in detail elsewhere.  

\section{Conclusions}
\label{sec:conc}
We have studied the properties of loosely bound composite dark matter, including its cosmology, aspects of its structure, and in particular the dynamics of its interactions with SM nuclei and atoms. We have examined interactions where the constituents in the composite are coupled directly to either SM nucleons or electrons. We have found that for loosely bound composites with an inter-constituent spacing scale $\Lambda_D < 1~{\rm MeV}$, it is possible for dark matter to scatter many times with nuclei, and for the number of these interactions to scale with nucleon number as $\sim A^4$, for constituent masses well in excess of the nuclear mass, $\overline{m}_d \gg m_A$. This has provided the first details of a composite dark matter model that would scatter many times in a large volume detector like DEAP-3600, with a larger-than-nuclear-sized cross-section and $\sim A^4$ scaling.

In addition, we have found some new dynamical regimes of composite interactions. Here there has been a preliminary investigation of loosely bound composites that, through a coupling to electrons, scatter softly many times with atoms in a single passage through a low threshold dark matter detector. The detectability of this kind of composite interaction, and in particular the response of a low threshold detector to many soft scatters in a single transit, will need to be addressed in further studies. In future work, it will also be interesting to investigate the detection of loosely bound composites for the scenario that the constituents' binding energy is less than a typical nuclear recoil.

\section*{Acknowledgements}

We thank Nirmal Raj, Mira Sheahan, and Shawn Westerdale for useful discussions. This work was supported by the Arthur B. McDonald Canadian Astroparticle Physics Research Institute, the Natural Sciences and Engineering Research Council of Canada (NSERC), and the Canada Foundation for Innovation. Research at Perimeter Institute is supported by the Government of Canada through the Department of Innovation, Science, and Economic Development, and by the Province of Ontario.
GM also acknowledges support from the UC office of the President through the UCI Chancellor's Advanced Postdoctoral Fellowship and the U.S. National Science Foundation under Grant PHY-2210283. JFA is supported by the U.S. Department of Energy under Contract DE-AC02-76SF00515. CVC was also generously supported by Washington University in St. Louis through the Edwin Thompson Jaynes Postdoctoral Fellowship.

\appendix

\section{Details of Composite-Nucleus Scattering}
\label{app:details}

Computing the differential cross section for a dark composite to scatter elastically with an SM nucleus requires knowledge of the density distribution of both the nucleus and the dark composite, and is best done using potential scattering theory. As in the standard case of WIMP-nucleus scattering, we assume that the potential felt by a pointlike dark matter particle passing through a nucleus is proportional to the density of the nucleus
\begin{align}
    V_A(r) = \frac{V_0}{\rho_0} \rho_A(r)\,.
\end{align}
We can write the potential between a composite dark matter state and a nucleus as a convolution of the dark matter number density with the potential sourced by the nucleus
\begin{align}
    V(\textbf{r}) &= \frac{1}{m_d}\int d^3r' V_A(\textbf{r})\rho_D(\textbf{r}-\textbf{r}')\\
    &=\frac{V_0}{m_d\rho_0}\int d^3r' \rho_A(\textbf{r})\rho_D(\textbf{r}-\textbf{r}')~.
\end{align}
In the weakly-coupled limit, $i.e.$ the limit in which the dark matter-nucleon interaction can be treated using the Born Approximation, the scattering amplitude can be written as

\begin{align}
    f &= \frac{-\mu_{AD}}{2\pi}\int d^3r e^{i \textbf{q}\cdot\textbf{r}} V(\textbf{r})\\
    &= -\sqrt{2\pi}\mu_{AD} \,\tilde{V}(\textbf{q})\,,
\end{align}
where $\mu_{AD}$ is the reduced mass of the composite-nucleus system and $\tilde{V}(\textbf{q})$ denotes the Fourier transform of $V(\textbf{r})$. 

In the case of a point dark matter particle scattering on a nucleus, the form factor of the nucleus encodes this momentum dependence, $i.e.$ $f \propto F(q)$. Assuming again that $V_A(\textbf{r}) \propto \rho_A(\textbf{r})$,

\begin{align}
    F_A(q) &= \frac{1}{m_A}\int d^3r e^{i \textbf{q}\cdot\textbf{r}} \rho_A(r)\\
    &= \frac{4\pi}{q m_A}\int dr r \sin(qr) \rho_A(r)\,,
\end{align}
where we have assumed that the density profile is spherically symmetric (see Ref.~\cite{Duda:2006uk} for details on commonly used nuclear form factors). We can define the dark composite form factor $F_D(q)$ in the same way, with the replacement $A \rightarrow D$ for the dark composite.

Next we treat the joint form-factor suppression for nucleus-composite scattering. In this case, the Convolution Theorem states that the Fourier transform of a convolution is just the product of the Fourier transforms, or more precisely

\begin{align}
    \tilde{V}(\textbf{q}) = \frac{V_0}{m_d\rho_0}(2\pi)^{3/2}\tilde{\rho}_A(\textbf{q})\tilde{\rho}_D(\textbf{q})\,,
\end{align}
where for either density,

\begin{align}
    \tilde{\rho_i}(\textbf{q}) &= \frac{1}{(2\pi)^{3/2}} \int d^3r e^{i \textbf{q}\cdot\textbf{r}} \rho_i(\textbf{r})\\
    &= \frac{\sqrt{2\pi}}{q} \int dr r \sin(qr) \rho_i(r)\\
    &=\frac{m_i}{2\sqrt{2\pi}}F_i(q)\,.
\end{align}
Putting all of this together,

\begin{align}
    f = -\frac{\pi \mu_{AD} V_0 m_A m_D}{2m_d\rho_0}F_A(q)F_D(q)\,
\end{align}
and thus
\begin{align}
    \frac{d\sigma}{d\Omega} = \frac{\pi^2 \mu_{AD}^2 V_0^2 m_A^2 m_D^2}{4m_d^2\rho_0^2}|F_A(q)|^2|F_D(q)|^2\,.
\end{align}

By analogy to all of the above, we can compute $\sigma_{nd}$, the zero-momentum-transfer cross section for dark matter (constituent)-nucleon scattering. So far, we have made no assumptions about the density profile of the dark composite: the above holds for, $e.g.$, any of the $N_D$ vs. $R_D$ scalings given in Ref.~\cite{Gresham:2017zqi}. If we now assume for concreteness that $\rho_d \simeq \rho_D$, $i.e.$ that the volume of a dark composite is approximately the volume of a constituent times the number of constituents, and thus $R_D \propto N_D^{1/3}$, as is the case for SM nuclei, we find

\begin{align}
    \sigma_{nd} = \frac{\pi^3 \mu_{nd}^2 V_0^2 m_n^2 m_d^2}{m_d^2 \rho_0^2}\,.
\end{align}
So finally, we obtain the differential cross section for elastic composite scattering, in terms of the zero-momentum constituent-nucleon cross-section

\begin{align}\label{eq:dmcoherent}
    \frac{d\sigma}{d\Omega} = \frac{1}{4\pi}\left(\frac{\mu_{AD}}{\mu_{nd}}\right)^2A^2N_D^2\sigma_{nd}F_A^2(q)F_D^2(q)\,.
\end{align}
The derivation above is approximately correct when the geometric size of the nucleus is similar to that of the dark composite. However, a correction must be made when this is not the case, due to the wave function of the nucleons or dark constituents being localized within the nucleus or dark composite, respectively. The scattering amplitude is modified by a factor of the ratio of the volumes of the two composites, such that the differential cross section is multiplied by a geometric factor $g^2$

\begin{align}
    g^2 = \textrm{min}\left[1,\,\left(\frac{L_{A}}{R_D}\right)^3\right]^2\,.
\end{align}
When the radius of the composite is very large compared to the inverse momentum transfer in scattering with a nucleus, incoherent scattering can dominate over coherent, elastic scattering. In this case, the total cross section to scatter with a nucleus is just $N_D$ times the constituent-nucleus cross-section

\begin{equation}\label{eq:incoherent}
    \frac{d\sigma_{inc}}{d\Omega} = \frac{1}{4\pi}\left(\frac{\mu_{Ad}}{\mu_{nd}}\right)^2A^2N_D\sigma_{nd}F_A^2(q)S_{D}(q)\,.
\end{equation}
Here, $S_D(q)$ is a structure function for the composite, which contains details about the binding energy of the constituents. When the binding energy is negligibly small, $S_D(q) \simeq 1$. Examining this equation, we see that when $m_d \gg m_A$, this cross section scales with $A^4$, but is not limited to the geometric size of the nucleus, but rather that of the composite.

\bibliography{apssamp}

\providecommand{\noopsort}[1]{}\providecommand{\singleletter}[1]{#1}%
\providecommand{\href}[2]{#2}\begingroup\raggedright\begin{thebibliography}{10}

\bibitem{Witten:1984rs}
E.~Witten, \emph{{Cosmic Separation of Phases}},
  \href{http://dx.doi.org/10.1103/PhysRevD.30.272}{\emph{Phys. Rev. D} {\bf 30}
  (1984) 272--285}.

\bibitem{Farhi:1984qu}
E.~Farhi and R.~Jaffe, \emph{{Strange Matter}},
  \href{http://dx.doi.org/10.1103/PhysRevD.30.2379}{\emph{Phys. Rev. D} {\bf
  30} (1984) 2379}.

\bibitem{DeRujula:1984axn}
A.~De~Rujula and S.~Glashow, \emph{{Nuclearites: A Novel Form of Cosmic
  Radiation}}, \href{http://dx.doi.org/10.1038/312734a0}{\emph{Nature} {\bf
  312} (1984) 734--737}.

\bibitem{Goodman:1984dc}
M.~W. Goodman and E.~Witten, \emph{{Detectability of Certain Dark Matter
  Candidates}}, \href{http://dx.doi.org/10.1103/PhysRevD.31.3059}{\emph{Phys.
  Rev. D} {\bf 31} (1985) 3059}.

\bibitem{Drukier:1986tm}
A.~Drukier, K.~Freese and D.~Spergel, \emph{{Detecting Cold Dark Matter
  Candidates}}, \href{http://dx.doi.org/10.1103/PhysRevD.33.3495}{\emph{Phys.
  Rev. D} {\bf 33} (1986) 3495--3508}.

\bibitem{Nussinov:1985xr}
S.~Nussinov, \emph{{Technocosmology: could a technibaryon excess provide a
  'natural' missing mass candidate?}},
  \href{http://dx.doi.org/10.1016/0370-2693(85)90689-6}{\emph{Phys. Lett. B}
  {\bf 165} (1985) 55--58}.

\bibitem{Bagnasco:1993st}
J.~Bagnasco, M.~Dine and S.~D. Thomas, \emph{{Detecting technibaryon dark
  matter}}, \href{http://dx.doi.org/10.1016/0370-2693(94)90830-3}{\emph{Phys.
  Lett. B} {\bf 320} (1994) 99--104},
  [\href{http://arxiv.org/abs/hep-ph/9310290}{{\tt hep-ph/9310290}}].

\bibitem{Alves:2009nf}
D.~S.~M. Alves, S.~R. Behbahani, P.~Schuster and J.~G. Wacker, \emph{{Composite
  Inelastic Dark Matter}},
  \href{http://dx.doi.org/10.1016/j.physletb.2010.08.006}{\emph{Phys. Lett.}
  {\bf B692} (2010) 323--326}, [\href{http://arxiv.org/abs/0903.3945}{{\tt
  0903.3945}}].

\bibitem{Kribs:2009fy}
G.~D. Kribs, T.~S. Roy, J.~Terning and K.~M. Zurek, \emph{{Quirky Composite
  Dark Matter}},
  \href{http://dx.doi.org/10.1103/PhysRevD.81.095001}{\emph{Phys. Rev. D} {\bf
  81} (2010) 095001}, [\href{http://arxiv.org/abs/0909.2034}{{\tt 0909.2034}}].

\bibitem{Lee:2013bua}
H.~M. Lee, M.~Park and V.~Sanz, \emph{{Gravity-mediated (or Composite) Dark
  Matter}}, \href{http://dx.doi.org/10.1140/epjc/s10052-014-2715-8}{\emph{Eur.
  Phys. J. C} {\bf 74} (2014) 2715},
  [\href{http://arxiv.org/abs/1306.4107}{{\tt 1306.4107}}].

\bibitem{Krnjaic:2014xza}
G.~Krnjaic and K.~Sigurdson, \emph{{Big Bang Darkleosynthesis}},
  \href{http://dx.doi.org/10.1016/j.physletb.2015.11.001}{\emph{Phys. Lett.}
  {\bf B751} (2015) 464--468}, [\href{http://arxiv.org/abs/1406.1171}{{\tt
  1406.1171}}].

\bibitem{Detmold:2014qqa}
W.~Detmold, M.~McCullough and A.~Pochinsky, \emph{{Dark Nuclei I: Cosmology and
  Indirect Detection}},
  \href{http://dx.doi.org/10.1103/PhysRevD.90.115013}{\emph{Phys. Rev.} {\bf
  D90} (2014) 115013}, [\href{http://arxiv.org/abs/1406.2276}{{\tt
  1406.2276}}].

\bibitem{Jacobs:2014yca}
D.~M. Jacobs, G.~D. Starkman and B.~W. Lynn, \emph{{Macro Dark Matter}},
  \href{http://dx.doi.org/10.1093/mnras/stv774}{\emph{Mon. Not. Roy. Astron.
  Soc.} {\bf 450} (2015) 3418--3430},
  [\href{http://arxiv.org/abs/1410.2236}{{\tt 1410.2236}}].

\bibitem{Wise:2014ola}
M.~B. Wise and Y.~Zhang, \emph{{Yukawa Bound States of a Large Number of
  Fermions}}, \href{http://dx.doi.org/10.1007/JHEP10(2015)165,
  10.1007/JHEP02(2015)023}{\emph{JHEP} {\bf 02} (2015) 023},
  [\href{http://arxiv.org/abs/1411.1772}{{\tt 1411.1772}}].

\bibitem{Hardy:2014mqa}
E.~Hardy, R.~Lasenby, J.~March-Russell and S.~M. West, \emph{{Big Bang
  Synthesis of Nuclear Dark Matter}},
  \href{http://dx.doi.org/10.1007/JHEP06(2015)011}{\emph{JHEP} {\bf 06} (2015)
  011}, [\href{http://arxiv.org/abs/1411.3739}{{\tt 1411.3739}}].

\bibitem{Hardy:2015boa}
E.~Hardy, R.~Lasenby, J.~March-Russell and S.~M. West, \emph{{Signatures of
  Large Composite Dark Matter States}},
  \href{http://dx.doi.org/10.1007/JHEP07(2015)133}{\emph{JHEP} {\bf 07} (2015)
  133}, [\href{http://arxiv.org/abs/1504.05419}{{\tt 1504.05419}}].

\bibitem{Bramante:2018qbc}
J.~Bramante, B.~Broerman, R.~F. Lang and N.~Raj, \emph{{Saturated Overburden
  Scattering and the Multiscatter Frontier: Discovering Dark Matter at the
  Planck Mass and Beyond}},
  \href{http://dx.doi.org/10.1103/PhysRevD.98.083516}{\emph{Phys. Rev.} {\bf
  D98} (2018) 083516}, [\href{http://arxiv.org/abs/1803.08044}{{\tt
  1803.08044}}].

\bibitem{Gresham:2018anj}
M.~I. Gresham, H.~K. Lou and K.~M. Zurek, \emph{{Astrophysical Signatures of
  Asymmetric Dark Matter Bound States}},
  \href{http://dx.doi.org/10.1103/PhysRevD.98.096001}{\emph{Phys. Rev.} {\bf
  D98} (2018) 096001}, [\href{http://arxiv.org/abs/1805.04512}{{\tt
  1805.04512}}].

\bibitem{Bramante:2018tos}
J.~Bramante, B.~Broerman, J.~Kumar, R.~F. Lang, M.~Pospelov and N.~Raj,
  \emph{{Foraging for dark matter in large volume liquid scintillator neutrino
  detectors with multiscatter events}},
  \href{http://dx.doi.org/10.1103/PhysRevD.99.083010}{\emph{Phys. Rev. D} {\bf
  99} (2019) 083010}, [\href{http://arxiv.org/abs/1812.09325}{{\tt
  1812.09325}}].

\bibitem{Ibe:2018juk}
M.~Ibe, A.~Kamada, S.~Kobayashi and W.~Nakano, \emph{{Composite Asymmetric Dark
  Matter with a Dark Photon Portal}},
  \href{http://dx.doi.org/10.1007/JHEP11(2018)203}{\emph{JHEP} {\bf 11} (2018)
  203}, [\href{http://arxiv.org/abs/1805.06876}{{\tt 1805.06876}}].

\bibitem{Grabowska:2018lnd}
D.~M. Grabowska, T.~Melia and S.~Rajendran, \emph{{Detecting Dark Blobs}},
  \href{http://dx.doi.org/10.1103/PhysRevD.98.115020}{\emph{Phys. Rev. D} {\bf
  98} (2018) 115020}, [\href{http://arxiv.org/abs/1807.03788}{{\tt
  1807.03788}}].

\bibitem{Coskuner:2018are}
A.~Coskuner, D.~M. Grabowska, S.~Knapen and K.~M. Zurek, \emph{{Direct
  Detection of Bound States of Asymmetric Dark Matter}},
  \href{http://dx.doi.org/10.1103/PhysRevD.100.035025}{\emph{Phys. Rev. D} {\bf
  100} (2019) 035025}, [\href{http://arxiv.org/abs/1812.07573}{{\tt
  1812.07573}}].

\bibitem{Bai:2018dxf}
Y.~Bai, A.~J. Long and S.~Lu, \emph{{Dark Quark Nuggets}},
  \href{http://dx.doi.org/10.1103/PhysRevD.99.055047}{\emph{Phys. Rev. D} {\bf
  99} (2019) 055047}, [\href{http://arxiv.org/abs/1810.04360}{{\tt
  1810.04360}}].

\bibitem{Digman:2019wdm}
M.~C. Digman, C.~V. Cappiello, J.~F. Beacom, C.~M. Hirata and A.~H. Peter,
  \emph{{Not as big as a barn: Upper bounds on dark matter-nucleus cross
  sections}}, \href{http://dx.doi.org/10.1103/PhysRevD.100.063013}{\emph{Phys.
  Rev. D} {\bf 100} (2019) 063013},
  [\href{http://arxiv.org/abs/1907.10618}{{\tt 1907.10618}}].

\bibitem{Bai:2019ogh}
Y.~Bai and J.~Berger, \emph{{Nucleus Capture by Macroscopic Dark Matter}},
  \href{http://dx.doi.org/10.1007/JHEP05(2020)160}{\emph{JHEP} {\bf 05} (2020)
  160}, [\href{http://arxiv.org/abs/1912.02813}{{\tt 1912.02813}}].

\bibitem{Bramante:2019yss}
J.~Bramante, J.~Kumar and N.~Raj, \emph{{Dark matter astrometry at underground
  detectors with multiscatter events}},
  \href{http://dx.doi.org/10.1103/PhysRevD.100.123016}{\emph{Phys. Rev. D} {\bf
  100} (2019) 123016}, [\href{http://arxiv.org/abs/1910.05380}{{\tt
  1910.05380}}].

\bibitem{Bhoonah:2020dzs}
A.~Bhoonah, J.~Bramante, S.~Schon and N.~Song, \emph{{Detecting Composite Dark
  Matter with Long Range and Contact Interactions in Gas Clouds}},
  \href{http://arxiv.org/abs/2010.07240}{{\tt 2010.07240}}.

\bibitem{Clark:2020mna}
M.~Clark, A.~Depoian, B.~Elshimy, A.~Kopec, R.~F. Lang and J.~Qin,
  \emph{{Direct Detection Limits on Heavy Dark Matter}},
  \href{http://arxiv.org/abs/2009.07909}{{\tt 2009.07909}}.

\bibitem{Gresham:2017zqi}
M.~I. Gresham, H.~K. Lou and K.~M. Zurek, \emph{{Nuclear Structure of Bound
  States of Asymmetric Dark Matter}},
  \href{http://dx.doi.org/10.1103/PhysRevD.96.096012}{\emph{Phys. Rev.} {\bf
  D96} (2017) 096012}, [\href{http://arxiv.org/abs/1707.02313}{{\tt
  1707.02313}}].

\bibitem{Gresham:2017cvl}
M.~I. Gresham, H.~K. Lou and K.~M. Zurek, \emph{{Early Universe synthesis of
  asymmetric dark matter nuggets}},
  \href{http://dx.doi.org/10.1103/PhysRevD.97.036003}{\emph{Phys. Rev.} {\bf
  D97} (2018) 036003}, [\href{http://arxiv.org/abs/1707.02316}{{\tt
  1707.02316}}].

\bibitem{Wise:2014jva}
M.~B. Wise and Y.~Zhang, \emph{{Stable Bound States of Asymmetric Dark
  Matter}}, \href{http://dx.doi.org/10.1103/PhysRevD.90.055030}{\emph{Phys.
  Rev. D} {\bf 90} (2014) 055030}, [\href{http://arxiv.org/abs/1407.4121}{{\tt
  1407.4121}}].

\bibitem{Acevedo:2020avd}
J.~F. Acevedo, J.~Bramante and A.~Goodman, \emph{{Nuclear fusion inside dark
  matter}}, \href{http://dx.doi.org/10.1103/PhysRevD.103.123022}{\emph{Phys.
  Rev. D} {\bf 103} (2021) 123022},
  [\href{http://arxiv.org/abs/2012.10998}{{\tt 2012.10998}}].

\bibitem{Acevedo:2021kly}
J.~F. Acevedo, J.~Bramante and A.~Goodman, \emph{{Accelerating composite dark
  matter discovery with nuclear recoils and the Migdal effect}},
  \href{http://dx.doi.org/10.1103/PhysRevD.105.023012}{\emph{Phys. Rev. D} {\bf
  105} (2022) 023012}, [\href{http://arxiv.org/abs/2108.10889}{{\tt
  2108.10889}}].

\bibitem{Fedderke:2024hfy}
M.~A. Fedderke, D.~E. Kaplan, A.~Mathur, S.~Rajendran and E.~H. Tanin,
  \emph{{Fireball antinucleosynthesis}},
  \href{http://dx.doi.org/10.1103/PhysRevD.109.123028}{\emph{Phys. Rev. D} {\bf
  109} (2024) 12}, [\href{http://arxiv.org/abs/2402.15581}{{\tt 2402.15581}}].

\bibitem{Petraki:2013wwa}
K.~Petraki and R.~R. Volkas, \emph{{Review of asymmetric dark matter}},
  \href{http://dx.doi.org/10.1142/S0217751X13300287}{\emph{Int. J. Mod. Phys.}
  {\bf A28} (2013) 1330028}, [\href{http://arxiv.org/abs/1305.4939}{{\tt
  1305.4939}}].

\bibitem{Zurek:2013wia}
K.~M. Zurek, \emph{{Asymmetric Dark Matter: Theories, Signatures, and
  Constraints}},
  \href{http://dx.doi.org/10.1016/j.physrep.2013.12.001}{\emph{Phys. Rept.}
  {\bf 537} (2014) 91--121}, [\href{http://arxiv.org/abs/1308.0338}{{\tt
  1308.0338}}].

\bibitem{Balan:2024cmq}
S.~Balan et~al., \emph{{Resonant or asymmetric: The status of sub-GeV dark
  matter}},  \href{http://arxiv.org/abs/2405.17548}{{\tt 2405.17548}}.

\bibitem{Aghanim:2018eyx}
{\scshape Planck} collaboration, N.~Aghanim et~al., \emph{{Planck 2018 results.
  VI. Cosmological parameters}},
  \href{http://dx.doi.org/10.1051/0004-6361/201833910}{\emph{Astron.
  Astrophys.} {\bf 641} (2020) A6},
  [\href{http://arxiv.org/abs/1807.06209}{{\tt 1807.06209}}].

\bibitem{Bramante:2017obj}
J.~Bramante and J.~Unwin, \emph{{Superheavy Thermal Dark Matter and Primordial
  Asymmetries}}, \href{http://dx.doi.org/10.1007/JHEP02(2017)119}{\emph{JHEP}
  {\bf 02} (2017) 119}, [\href{http://arxiv.org/abs/1701.05859}{{\tt
  1701.05859}}].

\bibitem{greiner2012nuclear}
W.~Greiner, D.~Bromley and J.~Maruhn, \emph{Nuclear Models}.
\newblock Springer Berlin Heidelberg, 2012.

\bibitem{DEAPCollaboration:2021raj}
{\scshape (DEAP Collaboration)\textdaggerdbl{}, DEAP} collaboration,
  P.~Adhikari et~al., \emph{{First Direct Detection Constraints on Planck-Scale
  Mass Dark Matter with Multiple-Scatter Signatures Using the DEAP-3600
  Detector}},
  \href{http://dx.doi.org/10.1103/PhysRevLett.128.011801}{\emph{Phys. Rev.
  Lett.} {\bf 128} (2022) 011801}, [\href{http://arxiv.org/abs/2108.09405}{{\tt
  2108.09405}}].

\bibitem{Kopp_2009}
J.~Kopp, V.~Niro, T.~Schwetz and J.~Zupan, \emph{Dama/libra data and
  leptonically interacting dark matter},
  \href{http://dx.doi.org/10.1103/physrevd.80.083502}{\emph{Physical Review D}
  {\bf 80} (Oct., 2009) }.

\bibitem{PhysRevD.85.076007}
R.~Essig, J.~Mardon and T.~Volansky, \emph{Direct detection of sub-gev dark
  matter}, \href{http://dx.doi.org/10.1103/PhysRevD.85.076007}{\emph{Phys. Rev.
  D} {\bf 85} (Apr, 2012) 076007}.

\bibitem{Bunge1993}
C.~Bunge, J.~Barrientos and A.~Vivier-bunge, \emph{Roothaan-hartree-fock
  ground-state atomic wave-functions - slater-type orbital expansions and
  expectation values for z=2-54},
  \href{http://dx.doi.org/10.1006/adnd.1993.1003}{\emph{Atom. Data Nucl. Data
  Tables} {\bf 53} (01, 1993) 113--162}.

\bibitem{Cappiello:2020lbk}
C.~V. Cappiello, J.~Collar and J.~F. Beacom, \emph{{New Experimental
  Constraints in a New Landscape for Composite Dark Matter}},
  \href{http://arxiv.org/abs/2008.10646}{{\tt 2008.10646}}.

\bibitem{Bramante:2022pmn}
J.~Bramante, J.~Kumar, G.~Mohlabeng, N.~Raj and N.~Song, \emph{{Light dark
  matter accumulating in planets: Nuclear scattering}},
  \href{http://dx.doi.org/10.1103/PhysRevD.108.063022}{\emph{Phys. Rev. D} {\bf
  108} (2023) 063022}, [\href{http://arxiv.org/abs/2210.01812}{{\tt
  2210.01812}}].

\bibitem{SBC:2021yal}
{\scshape SBC, CE\ensuremath{\nu}NS Theory Group at IF-UNAM} collaboration,
  L.~J. Flores et~al., \emph{{Physics reach of a low threshold scintillating
  argon bubble chamber in coherent elastic neutrino-nucleus scattering reactor
  experiments}},
  \href{http://dx.doi.org/10.1103/PhysRevD.103.L091301}{\emph{Phys. Rev. D}
  {\bf 103} (2021) L091301}, [\href{http://arxiv.org/abs/2101.08785}{{\tt
  2101.08785}}].

\bibitem{Kahn:2021ttr}
Y.~Kahn and T.~Lin, \emph{{Searches for light dark matter using condensed
  matter systems}},
  \href{http://dx.doi.org/10.1088/1361-6633/ac5f63}{\emph{Rept. Prog. Phys.}
  {\bf 85} (2022) 066901}, [\href{http://arxiv.org/abs/2108.03239}{{\tt
  2108.03239}}].

\bibitem{Duda:2006uk}
G.~Duda, A.~Kemper and P.~Gondolo, \emph{{Model Independent Form Factors for
  Spin Independent Neutralino-Nucleon Scattering from Elastic Electron
  Scattering Data}},
  \href{http://dx.doi.org/10.1088/1475-7516/2007/04/012}{\emph{JCAP} {\bf 04}
  (2007) 012}, [\href{http://arxiv.org/abs/hep-ph/0608035}{{\tt
  hep-ph/0608035}}].

\end{thebibliography}\endgroup
\bibliographystyle{jhep}

\end{document}